\documentclass[notitlepage]{revtex4-1} 

\usepackage{amsmath}  
\usepackage{amsfonts} 
\usepackage{graphicx} 
\usepackage{subcaption}
\usepackage{xcolor}

\begin{document}

\title{Exploring Hubble Constant Data in an Introductory Course}

\author{Jeffrey M. Hyde}
\email{jhyde@bowdoin.edu}
\affiliation{Department of Physics \& Astronomy, Bowdoin College, Brunswick, ME 04011}

\date{\today}

\maketitle 

\section{Introduction}\label{sec:intro}
Popular accounts of exciting discoveries often draw students to physics and astronomy, but at the introductory level it is challenging to connect with these in a meaningful way. The use of real astronomical data in the classroom can help bridge this gap and build valuable quantitative and scientific reasoning skills \cite{Oostra:2006,LoPresto:2019}. This paper presents a strategy for studying Hubble's Law and the accelerating expansion of the universe using actual data. Along with understanding the physical concepts, an explicit goal is to develop skills for analyzing data in terms of a model.

Hubble's Law is the observation that distant galaxies are receding from us at a rate $v$ proportional to their distance $D$ \cite{Hubble:1929ig}:
\begin{align}\label{eq:hubble_law}
v & = H D.
\end{align}
Knowledge of the Hubble ``constant" (or Hubble parameter) $H$ can help answer interesting questions, such as the age and matter content of the universe. Students may have read about recent debates over its present value (denoted $H_0$) \cite{overbye2017cosmos}. Many excellent resources are devoted to teaching the concept \cite{wallace:2011,wallace:2012,Prather:2013,Forringer:2014,Lincoln:2014,Belenkiy:2015}.

In order to incorporate analysis of real data in a meaningful way I use activities that develop two important skills. First, students use a tangible model of an expanding universe to find a value of $H_0$. This builds a working understanding of how the model relates to features that may be observed in data. Second, guided questions for three ``case studies" show students how to constructively deal with confusing points in the analysis of data. Before starting this activity, I have students complete a Lecture-Tutorial on the expansion of the universe \cite{Prather:2013}. This introduces students to basic ideas of Hubble plots and addresses some of the other conceptual difficulties related to the expansion of the universe.

\section{Finding $H_0$ in a Model Expanding Universe}\label{sec:model}

To model cosmological expansion, two sheets of paper present “snapshots” of some part of the universe at different times. Randomly placed dots represent galaxies, and the later snapshot is made by uniformly stretching the entire coordinate grid by some factor. Students compute the distances from a certain galaxy to several others, and do the same with the later snapshot.

The recession velocity is then computed from the change in distance over the time interval represented by the two sheets. After plotting recession velocity versus distance as seen from the galaxy, the slope will be $H_0$. I use the scale 1 cm = 1 Mpc ($10^6$ parsec), and stretch the distance grid by a multiplicative factor of 1.4 between snapshots 3 Gyr apart, where 1 Gyr = $10^9$ years. This leads to a value of $H_0 \approx 130$ km/s/Mpc $\approx 0.4$ Gyr$^{-1}$. It's useful to note that stretching distances by a factor of $m$ over $T$ Gyr leads to $H \approx \frac{m-1}{T}\times 10^3$ km/s/Mpc, using 1 Mpc/Gyr = 980 km/s $\approx 1000$ km/s.

To emphasize that this is a property of the expansion and not one specific galaxy, we compute $H_0$ as observed from another galaxy in the example universe, which should give the same value. Follow-up questions build on this tangible model and prepare students for later interpretation of data. Useful questions directly relevant to the next section are “What would change if $H_0$ were larger or smaller?” and “would a larger $H_0$ correspond to an older or younger universe today?” Observationally-oriented what-if’s are also important. For instance, ``are there any factors that the model does not take into account? How would these affect the plot?"

A few possible points of confusion are worth mentioning. First, connecting recession velocity to the change in distance using $v = \Delta x / \Delta t$ can be conceptually difficult. Second, since each velocity comes from comparison of two distances, what is ``the" distance? This ambiguity can be minimized by shortening the time interval in between snapshots and correspondingly reducing the multiplicative factor $m$. However, too small of a time interval can make it difficult to reliably measure the smallest distances and therefore obscure the linear relationship we are seeking to demonstrate. Factors such as the increments on rulers used by students and the separation of galaxies on the page influence the best choice of parameters, so it is worthwhile to experiment with some different values.

\section{Analyzing Data - Three Examples}\label{sec:analysis}

After the hands-on experience described above, students are ready to examine three “case studies” to sharpen their thinking about models and their relation to data. It will be important to know that distances are determined by measuring the apparent brightness of ``standard candles" - stars or supernovae with a known absolute magnitude. Depending on factors such as time constraints, students could plot the data themselves using tables in the cited papers, or simply be given the plots.

\subsection{Hubble's Original Data}

\begin{figure}[t]
	\centering
	\begin{subfigure}[b]{0.45\textwidth}
		\centering
		\includegraphics[width=\textwidth]{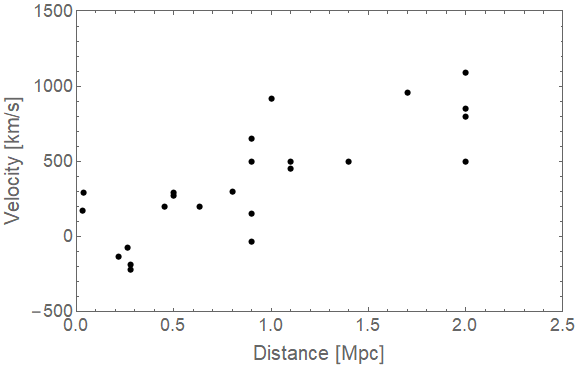}
		\caption{\label{fig:hubble1929}}
	\end{subfigure} \, \, \, \, \, \, %
	\begin{subfigure}[b]{0.45\textwidth}
		\centering
		\includegraphics[width=\textwidth]{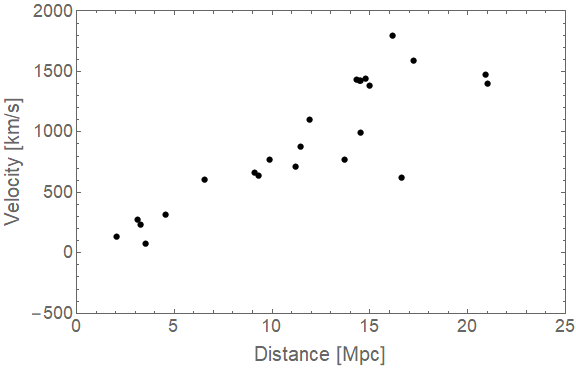}
		\caption{\label{fig:hubble_key_cepheids}}
	\end{subfigure}
\caption{\label{fig:hubble_comparison}Hubble's original data \cite{Hubble:1929ig} (Fig.~\ref{fig:hubble1929}) and galaxies whose distances were also determined using Cepheid variables as part of the Hubble Key Project \cite{Freedman:2000cf} (Fig.~\ref{fig:hubble_key_cepheids}).}
\end{figure}

Hubble's original data set \cite{Hubble:1929ig} gives an expansion rate of around 500 km/s/Mpc (Fig.~\ref{fig:hubble1929}). The data follow a linear trend, but with quite a bit of scatter - an opportunity for students to brainstorm possible explanations. In this case, all galaxies have their own ``peculiar motion" in addition to the Hubble expansion (stretching of the grid), so nearby galaxies may be moving towards us faster than expansion is taking them away. This is a good place to refer back to the earlier question of what factors the model did not take into account.

Hubble's value of $H_0$ implied the age of the universe was around 2 billion years, in contrast with early radiometric dating of rocks on Earth to 3-4 billion years \cite{Trimble:1996jqa}. How could the Earth be older than the universe? This question provides good motivation to think broadly about sources of error. We now know that Hubble's data had a systematic error where distances were underestimated (largely because two different types of Cepheid variable stars were assumed to be identical standard candles), although I defer this discussion with the class until after the following example.

\subsection{Hubble Key Project to measure $H_0$}

The Hubble Key Project compiled a variety of Hubble Space Telescope observations with the goal of decreasing uncertainties in $H_0$ \cite{Freedman:2000cf}. Fig.~\ref{fig:hubble1929} shows the result from their observation of Cepheid variables. This leads to an important question: is the data set shown in Fig.~\ref{fig:hubble1929} consistent with the data set shown in Fig.~\ref{fig:hubble_key_cepheids}?

Students may find it difficult to agree on an answer to ``what does it mean for two data sets to be consistent?" They often suggest that if the best-fit lines have any difference at all (no matter how small) then the data sets are totally inconsistent. This competes with another popular suggestion, that the spread of data points and uncertainty in slope renders the question impossible to answer. These responses are similar to misconceptions identified in introductory physics students' reasoning about data \cite{Buffler:2001}. In particular, the suggestion that any difference in measurements must mean they are inconsistent may be thought of as ``point paradigm" reasoning that supposes single measurements can represent true values with no inherent uncertainty.

\begin{figure}[t]
	\centering
	\begin{subfigure}[b]{0.45\textwidth}
		\centering
		\includegraphics[width=\textwidth]{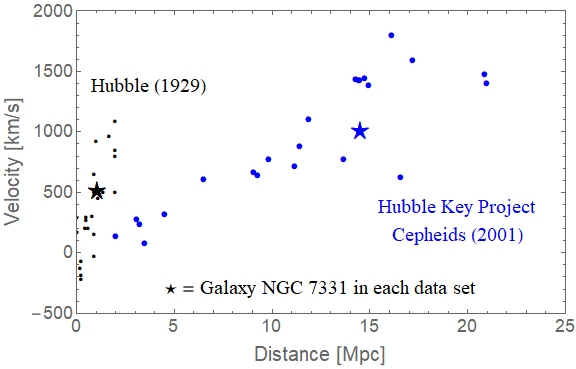}
		\caption{\label{fig:hubble_comparison_2}}
	\end{subfigure} \, \, \, \, \, \, %
	\begin{subfigure}[b]{0.45\textwidth}
		\centering
		\includegraphics[width=\textwidth]{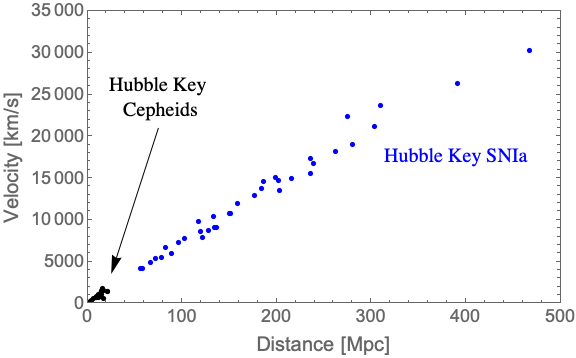}
		\caption{\label{fig:hubble_key_2methods}}
	\end{subfigure}
\caption{\label{fig:hubble_key}Fig.~\ref{fig:hubble_comparison_2} shows both data sets of Fig.~\ref{fig:hubble_comparison} plotted on the same axes. The galaxy NGC 7331 is present in both surveys, so it is marked with a star for comparison. Hubble systematically underestimated distances and therefore overestimated the slope, or Hubble parameter, $H_0$. Fig.~\ref{fig:hubble_key_2methods} shows the Hubble Key Project's measurement of $H_0$ using Cepheids (same data plotted in Fig.~\ref{fig:hubble_comparison_2}) and Type Ia supernovae \cite{Freedman:2000cf}.}
\end{figure}

If students are fitting the data with software, they could consider the quoted uncertainty in slope. A qualitative approach is to think about the range of slopes that could plausibly describe each data set. After plotting both on the same axes (see Fig.~\ref{fig:hubble_comparison_2}), it becomes clear that Hubble Key found a much lower value of $H_0$. Students should be able to conclude that this lower $H_0$ predicts an older universe than Hubble's original work, although this is often one of the questions that students find more difficult.

This is a good time to return to hypotheses about sources of error in Fig.~\ref{fig:hubble1929}, and bring up some of the historical context (e.g. measuring two populations of Cepheids) mentioned earlier. The galaxy NGC 7331 was part of both surveys, and it is interesting to compare its position in each (marked with a star in Fig.~\ref{fig:hubble_comparison_2}). The distance changes significantly, but the speed has also been corrected for local motions of our solar system, galaxy, and local group. Since different methods may have different systematic errors, it is also worthwhile to discuss results that come from a different standard candle. Fig.~\ref{fig:hubble_key_2methods} shows Type Ia supernovae giving about the same $H_0$ as Cepheids.

\subsection{Accelerated Expansion}

Studying accelerated expansion connects students directly with a discovery recognized by the 2011 Nobel Prize in physics \cite{Riess:1998cb,Perlmutter:1998np}, the relevant plots being Fig. 4 or 5 of \cite{Riess:1998cb}, or Fig. 1 of \cite{Perlmutter:1998np}. The benefits of using a figure from the published paper are: (1) the authenticity of the ``real thing" is exciting and (2) it has curves showing predictions of various models. Computing such curves is beyond the scope of an introductory class (see for example Chapter 6 of \cite{Ryden:2003yy}, along with the text of the papers cited here), but by this point students will be ready to understand the result at a qualitative level.

\begin{figure}[t]
	\centering
	\begin{subfigure}[b]{0.45\textwidth}
		\centering
		\includegraphics[width=\textwidth]{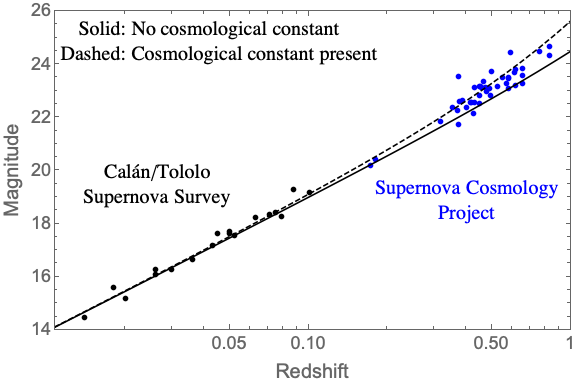}
	\end{subfigure}
\caption{\label{fig:accel_data}Standard candles at even greater distances reveal the accelerating expansion. This plot shows data from \cite{Perlmutter:1998np}, as well as two curves showing predictions based on the contents of the universe. The solid curve is a prediction if the energy density of the universe is all non-relativistic matter. The dashed curve is a prediction based on the now-accepted values for the energy density of the universe, where $70\%$ is cosmological constant or dark energy, and only $30\%$ is matter.}
\end{figure}

Fig.~\ref{fig:accel_data} shows a plot of the data from the Cal\'an/Tololo Supernova Survey \cite{Hamuy:1996sq} and the Supernova Cosmology Project \cite{Perlmutter:1998np}, along with two curves that show predictions based on the contents of the universe. The two predictions shown are for the case where our universe is comprised only of non-relativistic matter, and for the case where our universe is $30\%$ non-relativistic matter and $70\%$ cosmological constant (or dark energy). In Fig. 1 of \cite{Perlmutter:1998np}, the curve labeled $(\Omega_M, \Omega_{\Lambda}) = (1, 0)$ shows a ``traditional" case with no cosmological constant/dark energy, and comparison with the data points and error bars will convince students that the data points tend to fall above that line. These figures plot apparent magnitude versus redshift, so recession velocity is now on the horizontal axis. Guidance from the instructor can help students determine that the apparent magnitude of a standard candle is a measurement of distance.

For interpretation at a qualitative level, it helps to ask questions that point out specific pieces of reasoning. For instance, the earlier question about ``which galaxies emitted their light further in the past?" can be followed up with ``what would the Hubble plot look like if the universe was expanding faster in the past? Slower in the past?" A common mistake here is the assumption that time since the Big Bang increases from left to right on any graph.

\section{Further Exploration}

There are many possibilities for further exploration. For example, students can be asked to find data in the NASA Extragalactic Database (ned.ipac.caltech.edu) and make their own plots. Interesting discussions arise when they have to choose which of several given measurements is ``the" value they will use, or compare results after choosing samples with very different distance ranges or numbers of galaxies. Another possibility is to use data from the Sternberg Supernova Catalog, where interstellar extinction is seen to dim Type Ia supernovae \cite{Tsvetkov:2004}. Since this dimming increases with distance, students will have to determine that the closest supernovae provide the most reliable estimate of $H_0$.

With the guidance described in this paper, I've found this to be an accessible activity that challenges introductory students to think hard about Hubble's Law and how analysis of data is used to evaluate models.

\begin{acknowledgments}

I would like to thank the anonymous referees for valuable comments that improved the quality and clarity of this paper.

\end{acknowledgments}

\end{document}